\documentclass{PoS}

\usepackage{graphicx}
\usepackage{subfigure}
\usepackage{amsmath}
\usepackage{multirow}
\usepackage{mathrsfs}
\usepackage{slashed}

\title{$K_L-K_S$ mass difference from Lattice QCD}

\ShortTitle{$K_L-K_S$ mass difference from Lattice QCD}

\author{\speaker{Jianglei Yu}\\
				Physics Department, Columbia University, New York \\
				E-mail: \email{jy2379@columbia.edu}}

\abstract{We will report on the first full calculation of the $K_L-K_S$ mass difference in lattice QCD. The calculation is performed on a 2+1 flavor, domain wall fermion, 243×64 ensemble with a 329 MeV pion mass and a 575 MeV kaon mass. Both double penguin diagrams and disconnected diagrams are included in this calculation. The calculation is made finite through the GIM mechanism by introducing a 949 MeV valence charm quark. While the double penguin diagrams contribute a very small fraction to the mass difference, there is a large cancellation between disconnected diagrams and other types of diagrams. We obtain the mass difference $\Delta M_K$=$3.30(34)\times 10^{-12}$ MeV for these unphysical kinematics.}

\FullConference{31st International Symposium on Lattice Field Theory - LATTICE 2013 \\
		July 29 - August 3, 2013\\
		Mainz, Germany}

\begin{document}

\section{Introduction}
The kaon mass difference $\Delta M_K$ with a value of $3.483(6) \times 10^{-12}$ MeV~\cite{Nakamura:2010zzi} led to the prediction of charm quark fifty years ago. This extremely small mass difference is believed to arise from $K^0$-$\overline{K}^0$ mixing via second-order weak interaction. However, because it arises from an amplitude in which strangeness changes by two units, this is a promising quantity to reveal new phenomena which lie outside the standard model. In perturbation theory calculation, the standard model contribution to $\Delta M_K$ is separately into short distance and long distance parts. The short distance part receives most contributions from momenta on the order of the charm quark mass. As pointed out in the recent NNLO calculation~\cite{Brod:2011ty}, the NNLO terms are as large as 36\% of the leading order (LO) and next-to-leading order (NLO) terms, raising doubts about the convergence of QCD perturbation series at this energy scale. As for the long distance part of $\Delta M_K$, so far there is no result with controlled uncertainty available since it is highly non-perturbative. However, an estimation given by Donoghue {\it et al.}~\cite{Donoghue:1983hi} suggest that there can be sizable long distance contributions. 

Lattice QCD provides a fist-principle method to compute non-perturbative QCD effects in electroweak process. We have proposed a lattice method to compute $\Delta M_K$~\cite{Christ:2010zz,Christ:2012np}. Preliminary numerical works~\cite{Christ:2012se} have been done for $\Delta M_K$ on a 2+1 flavor $16^3\times 32$ DWF ensemble with a 421 MeV pion mass. We obtain a mass difference $\Delta M_K$ which ranges from $6.58(30)\times 10^{-12}$ MeV to $11.89(81) \times 10^{-12}$ MeV for kaon masses varying from 563 MeV to 839 MeV. The preliminary work only include parts of the diagrams, which means it is a non-unitary calculation. In this proceeding, we will report on a full calculation with a lighter pion mass including the effects of disconnected diagrams.

\section{Evaluation of $\Delta M_K$}
\begin{figure}[htp!]
	\centering
	\includegraphics[width=0.6\textwidth]{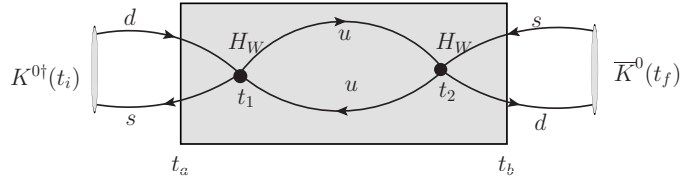}
	\caption{One type of diagram contributing to integrated correlator $\mathscr{A}$. Here $t_2$ and $t_1$ are integrated over the time interval $[t_a,t_b]$, represented by the shaded region.}
	\label{fig:int_correlator}
\end{figure}

We will briefly summarize the lattice method for evaluating $\Delta M_K$ here. More details can be found in~\cite{Christ:2012se}. The essential step is to perform a second-order integration of the product of two first-order weak Hamiltonians in a given space-time volume. 
\begin{equation}
\mathscr{A}=\frac{1}{2}\sum_{t_2=t_a}^{t_b}\sum_{t_1=t_a}^{t_b}\langle0|T\left\{\overline{K^0}(t_f)H_W(t_2)H_W(t_1)\overline{K^0}(t_i)\right\}|0\rangle.
\label{eq:integrated_correlator}
\end{equation}
This integrated correlator is represented schematically in Fig.~\ref{fig:int_correlator}. 
After inserting a sum over intermediate states and summing explicitly over $t_2$ and $t_1$ in the interval $[t_a,t_b]$ one obtains :
\begin{equation}
		\mathscr{A} =  N_K^2e^{-M_K(t_f-t_i)} \sum_{n}\frac{\langle\overline{K}^0|H_W|n\rangle\langle n|H_W|K^0\rangle}{M_K-E_n}\left( -T - \frac{1}{M_K-E_n} + \frac{e^{(M_K-E_n)T}}{M_K-E_n}\right).
\label{eq:integration_result}
\end{equation}
Here $T=t_b-t_a+1$ is the the interaction range. The coefficient of the term which is proportional to $T$ in Eq.~\eqref{eq:integration_result} gives us $\Delta M_K$ up to some renormalization factors :
\begin{equation}
\Delta M_K =  2 \sum_{n} \frac{\langle\overline{K}^0|H_W|n\rangle\langle n|H_W|K^0\rangle}{M_K-E_n}
\label{eq:massdiff}
\end{equation}
The exponential terms coming from states $|n\rangle$ with $E_n>M_K$ in Eq.~\eqref{eq:integration_result} are exponentially decreasing as $T$ increases. These terms are negligible for sufficiently large $T$. There will be exponentially increasing terms coming from $\pi^0$ and vacuum intermediate states. We evaluate the matrix element $\langle \pi^0 | H_W | K^0 \rangle$ and subtract the $\pi^0$ exponentially increasing term explicitly from Eq.~\eqref{eq:integration_result}. For the vacuum state, we add a pseudo-scalar density term to the weak Hamiltonian to eliminate the matrix element $\langle 0 | H_W + c_s \bar{s} \gamma^5 d| K^0 \rangle$. Since the pseudo-scalar density can be written as the divergence of the axial currents, the final mass difference will not be changed by adding this term. After the subtraction of exponentially increasing terms, a linear fit at sufficiently large $T$ will give us $\Delta M_K$.

\begin{figure}[!htp]
	\centering
\begin{tabular}{cc}
	\hline
\includegraphics[width=0.3\textwidth]{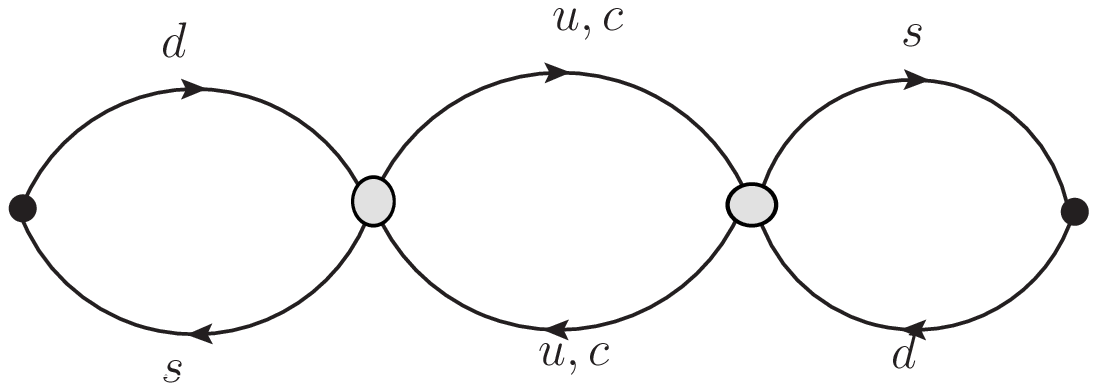} & \includegraphics[width=0.3\textwidth]{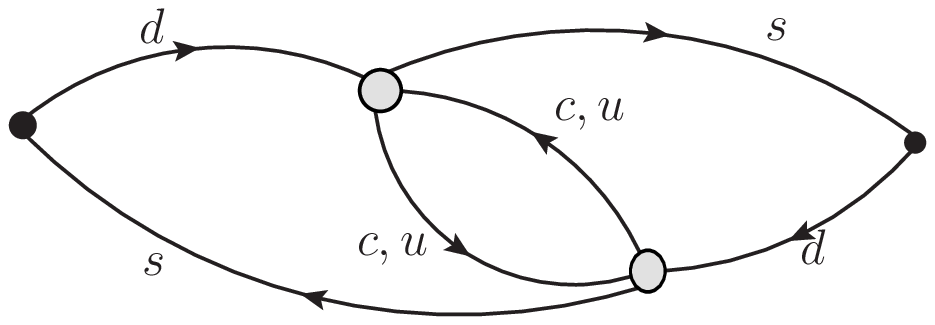} \\
Type 1 & Type 2 \\
\hline
\includegraphics[width=0.3\textwidth]{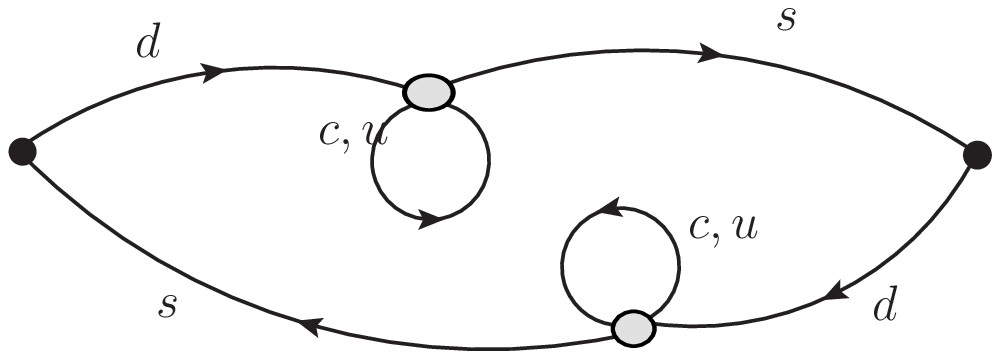} & \includegraphics[width=0.3\textwidth]{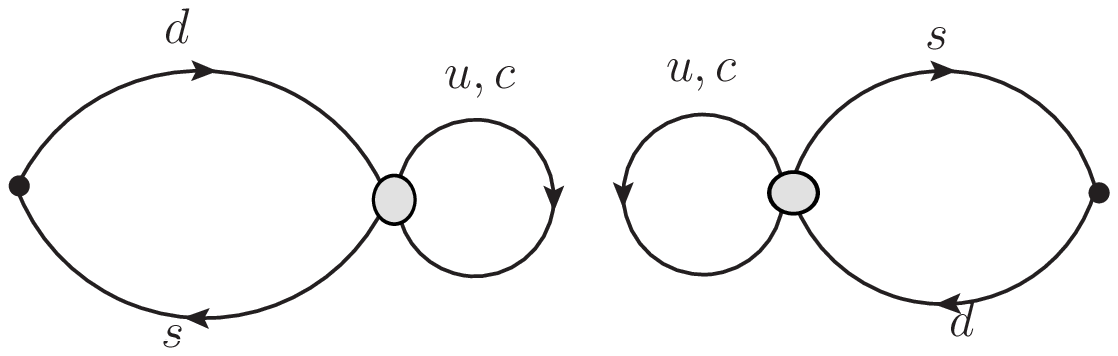}\\
Type 3 & Type 4\\
\hline
\end{tabular}
\caption{Four type of diagrams contributing the mass difference calculation. The shaded circles represent $\Delta S=1$ four quark operators. The black dots are $\gamma_5$ insertions for kaon sources.}
\label{fig:diagrams}
\end{figure}

The $\Delta S=1$ effective Hamiltonian in this calculation is
\begin{equation}
H_W=\frac{G_F}{\sqrt{2}}\sum_{q,q^{\prime}=u,c}V_{qd}V^{*}_{q^{\prime}s}(C_1Q_1^{qq^{\prime}}+C_2Q_2^{qq^{\prime}})
\label{eq:H_W}
\end{equation}
where $V_{qd}$ and $V_{q's}$ are Cabibbo-Kobayashi-Maskawa (CKM) matrix elements, $C_1$ and $C_2$ are Wilson coefficients for the current-current operators, which are defined as:
\begin{equation}
\begin{split}
Q_1^{qq{\prime}}&=(\bar{s}_id_i)_{L}(\bar{q}_jq^{\prime}_j)_{L}\\
Q_2^{qq{\prime}}&=(\bar{s}_id_j)_{L}(\bar{q}_jq^{\prime}_i)_{L}\,,
\end{split}
\label{eq:operator}
\end{equation}
The Wilson coefficients are calculated in the $\overline{MS}$ scheme using NLO perpetuation theory~\cite{Buchalla:1995vs}. Then the $\overline{MS}$ operators and the lattice operators are connected by using a a Rome-Southampton style style non-perturbative renormalization method~\cite{Martinelli:1994ty}. Inserting the weak Hamiltonian into the four point correlators, there will be four type of diagrams as shown in Fig.~\ref{fig:diagrams}. In our previous work~\cite{Christ:2012se}, we include only first two types of diagrams. All the diagrams are included in this work. The type four diagrams, which are disconnected, are expected to be the main source of statistical noise.

\section{Details of simulation}
This calculation if performed on a lattice ensemble generated with the Iwasaki gauge action and 2+1 flavors of domain wall fermion. The space time volume is $24^3\times64$ and the inverse lattice spacing $a^{-1}=1.729(28)$ GeV. The fifth-dimensional extent is $L_s=16$ and the residual mass is $m_{res}=0.00308(4)$ in lattice units. The sea light quark and strange quark masses are $m_l=0.005$ and $m_s=0.04$, corresponding to a pion mass $M_{\pi}=330$ MeV and a kaon mass $M_{K}=575$ MeV. A valence charm quark with mass $m_c^{\overline{MS}}$(2 GeV) = 949 MeV is used to implement GIM cancellation. We use 800 configurations, each separated by 10 time units.

We will use Fig.~\ref{fig:int_correlator} to explain the set up of this calculation. We use Coulomb gauge fixed wall sources for the kaons. The two kaons are separated by $31$ in lattice unites. The two weak Hamiltonians are at least $6$ time slices away from the kaon sources so that the kaon interpolating operators can project onto kaon states. For type 1 and type 2 diagrams, we use the same strategy as in~\cite{Christ:2012se}. We compute a point source propagator on each time slice to  calculate the quark lines connecting the two weak Hamiltonians. For type 3 and type 4 diagrams, we calculate random wall source propagators to evaluate the quark loops. In order to reduce the noise coming from random numbers, we use $6$ sets of random number on each time slice. All the diagrams are averaged over all time translations to increase statistics. For the light quark propagators, which is the most expensive part of this calculation, we calculate the lowest 300 eigenvectors of the Dirac operator and use low mode deflation to accelerate the light quark inverters.  

\section{Fitting results}
\begin{figure}[!htp]
	\centering
	\subfigure[Integrated correlator]{
		\includegraphics[width=0.4\textwidth]{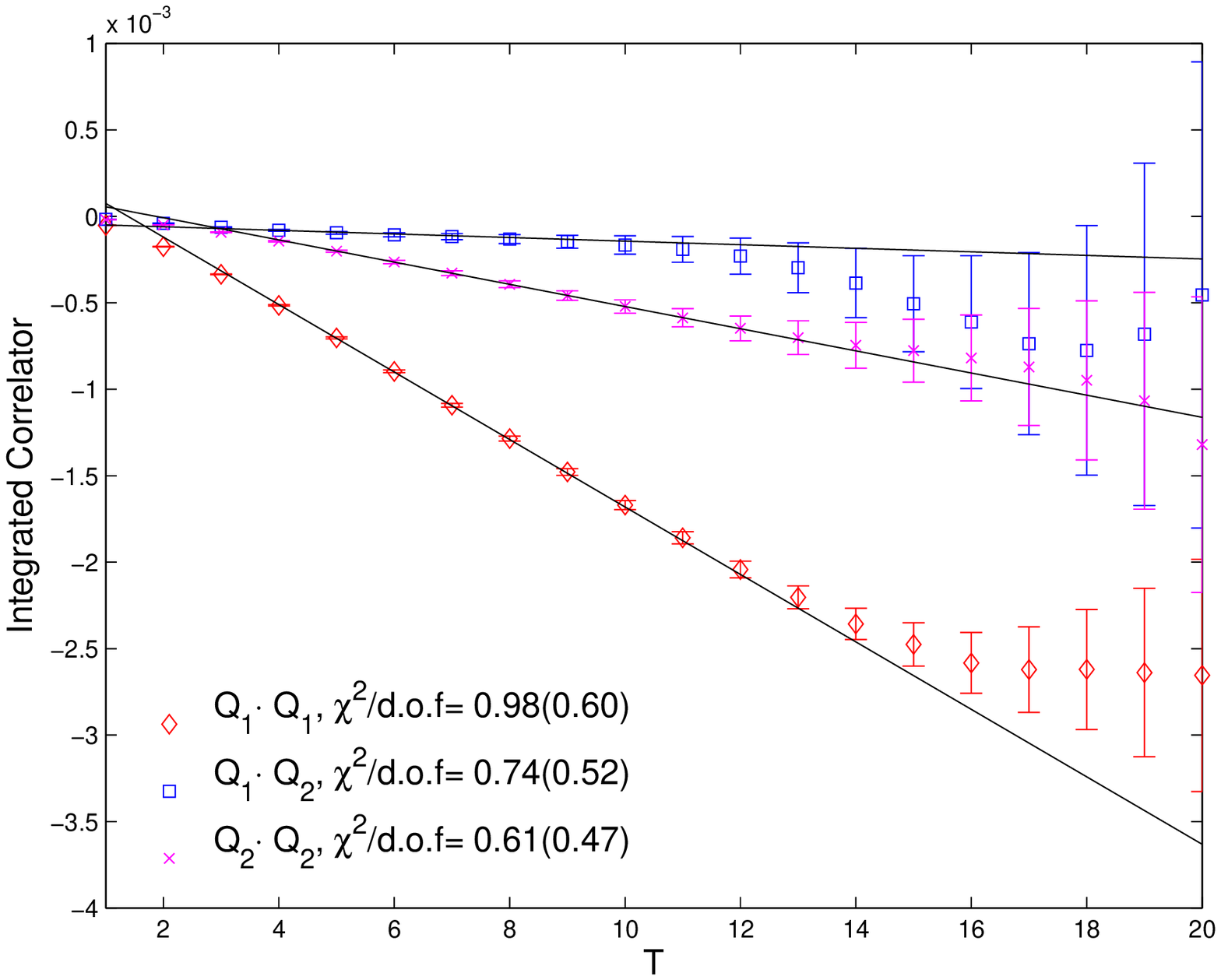} 
	\label{fig:int_corr}
}
\subfigure[Effective slope]{
		\includegraphics[width=0.4\textwidth]{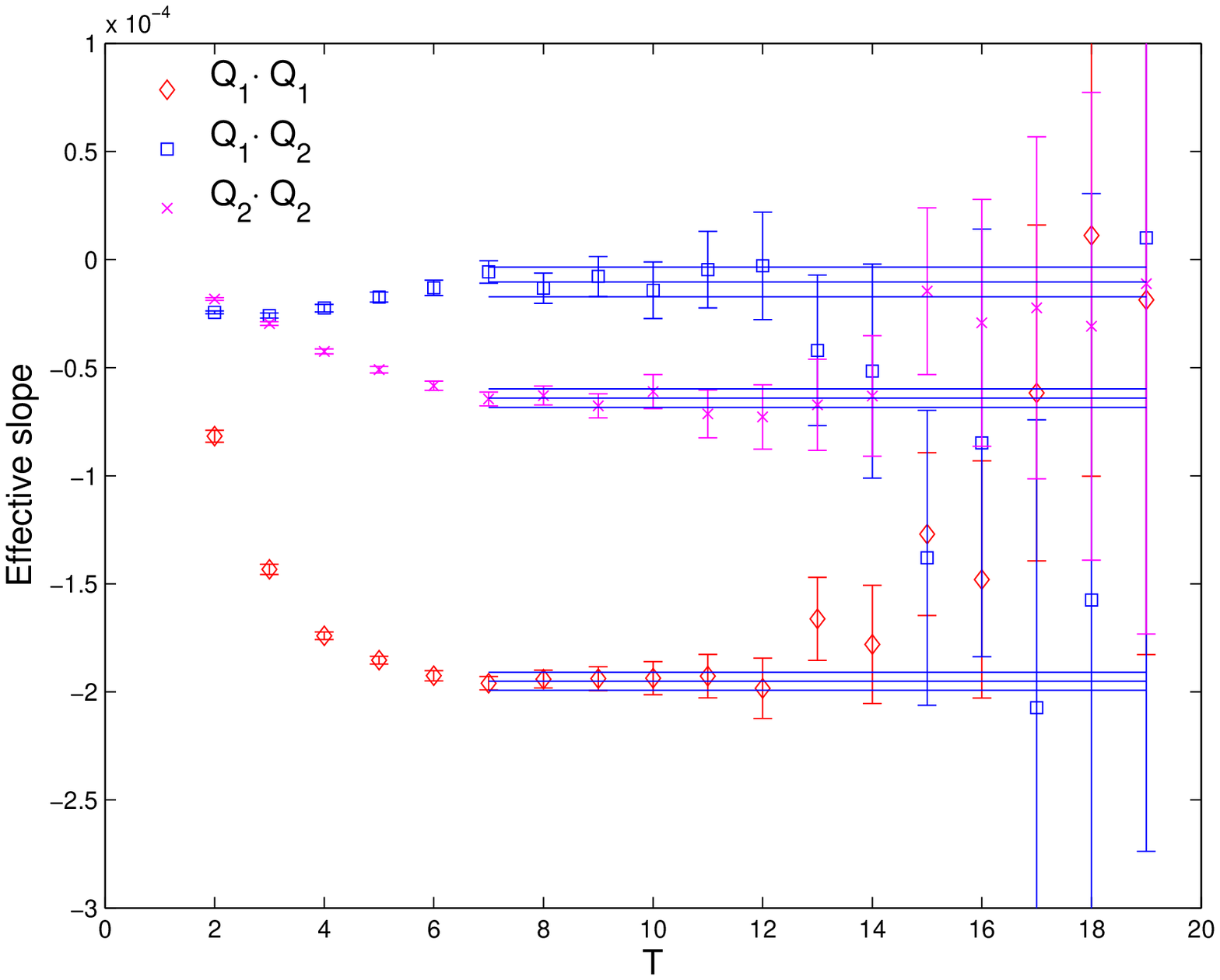} 
	\label{fig:effslope_v2}
}
	\caption{The left plot gives the integrated correlators for the three operator products $Q_1\cdot Q_1$, $Q_1\cdot Q_2$ and $Q_2\cdot Q_2$. The three lines give the linear fits to the data in the time interval [7,20]. The right plot gives the effective slope plots for three operator products.}
	\label{fig:all_diagrams}
\end{figure}

The results for the integrated correlators are given in Fig.~\ref{fig:int_corr}. Three curves correspond to three different operator combinations: $Q_1\cdot Q_1$, $Q_1\cdot Q_2$ and $Q_2\cdot Q_2$, respectively. The numbers are bare lattice results without any Wilson coefficients or renormalization factors. All the exponential increasing terms have been removed from the correlators. So we expect a linear behavior for large enough $T$. While $T$ becomes too large, the errors blow up. This is within our expectation since disconnected diagrams have exponentially increasing signal to noise ratio. The straight lines are the linear fitting results from the data points in the range $[7,20]$. The $\chi^2/d.o.f$ given in the figure suggest that these fits are robust.

Another method to check the quality of these fits are the effective slope plots, which is an analogy of the effective mass plots. The effective slope at a given time $T$ is calculated using a correlated fit with three data points at $T-1$, $T$ and $T+1$. In Fig.~\ref{fig:effslope_v2} we give the effective slope plots for three different operator combinations. The final fitting results and the errors are also given there. For operator combinations $Q_1\cdot Q_1$ and $Q_2\cdot Q_2$, we get good plateaus starting from $T=7$. The result for $Q_1\cdot Q_2$ is not so satisfying due to large error. However, as we will see later, the $Q_1\cdot Q_2$ contribution to $\Delta M_K$ is very small due to its small lattice amplitudes and its small Wilson coefficients. 

We have also tried different fittings to make sure that our results are not sensitive to the parameters we chose. There are two parameters we try to vary: the staring fitting time $T_{min}$ and the minimal separation between kaon sources and weak Hamiltonians $\Delta_{min}$. We first fix $\Delta_K=6$ and vary $T_{min}$ from 7 to 9. The result are given in Table.~\ref{tab:fit_tmin}. All the masses are in units of $10^{-12}$ MeV. While the central value of the fitting results are quite stable, the errors are very sensitive to the choice of $T_{min}$, which is a feature of disconnected diagrams. In Table.~\ref{tab:fit_deltak}, we give the results with fixing $T_{min}=7$ and $\Delta M_K$ from 6 to 8. Both the central values and the errors are very stable, suggesting that a separation of 6 is large enough to suppress the excited kaon states.

\begin{table}[!htp]
	\caption{The fitting results of mass difference for difference choice of $T_{min}$ while fixing $\Delta_K=6$. All the masses here are in units of $10^{-12}$ MeV.}
	\label{tab:fit_tmin}
	\centering
		\begin{tabular}{cccccc}
			\hline
			$\Delta_K$ & $T_{min}$ & $Q_1\cdot Q_1$ & $Q_1 \cdot Q_2$ & $Q_2 \cdot Q_2$ & $\Delta M_K$ \\
			\hline
			\multirow{3}{*}{6}	&	7	& 0.754(42)& -0.16(15)& 2.70(18)& 3.30(34) \\
		& 8  & 0.755(45) &-0.10(17)& 2.83(23)& 3.49(40) \\
	 &	9	& 0.758(53)& -0.16(22) &2.69(33)& 3.28(55) \\
			\hline
 \end{tabular}
\end{table}

\begin{table}[!htp]
	\caption{The fitting results of mass difference for difference choice of $\Delta_K$ while fixing $T_{min}=7$. All the masses here are in units of $10^{-12}$ MeV.}
	\label{tab:fit_deltak}
	\centering
		\begin{tabular}{cccccc}
			\hline
			$T_{min}$ & $\Delta_K$ & $Q_1\cdot Q_1$ & $Q_1 \cdot Q_2$ & $Q_2 \cdot Q_2$ & $\Delta M_K$ \\
			\hline
			\multirow{3}{*}{7}		&	6	& 0.754(42)& -0.16(15)& 2.70(18)& 3.30(34)\\
		&	7	& 0.755(42)& -0.18(15)& 2.66(18)& 3.23(34)\\
	 &	8	& 0.751(42)& -0.18(15)& 2.62(19)& 3.18(35)\\
			\hline
 \end{tabular}
\end{table}

\begin{figure}[!htp]
	\centering
	\subfigure[Integrated correlator]{
\includegraphics[width=0.4\textwidth]{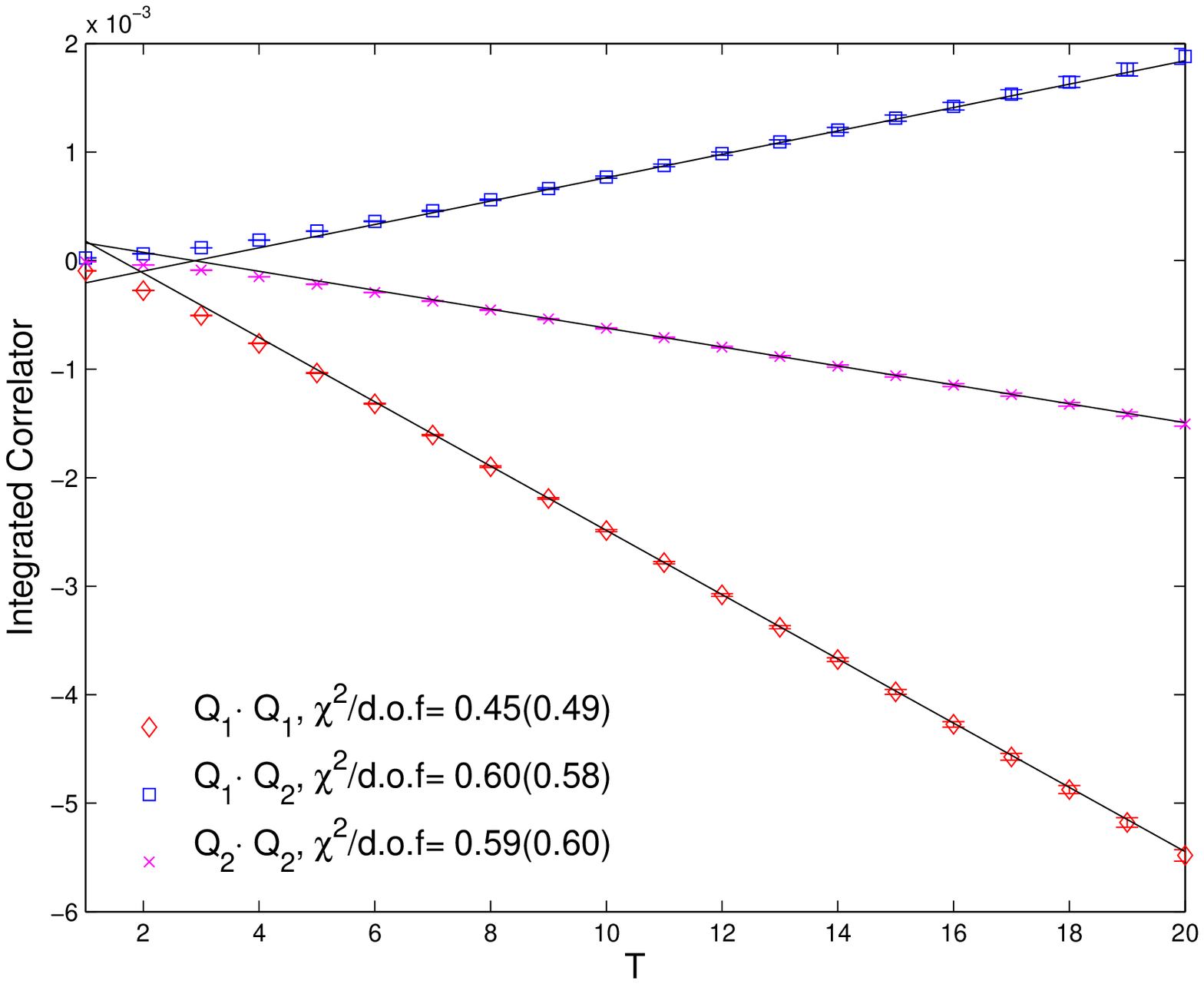} 
	\label{fig:int_corr_type12}
}
\subfigure[Effective slope]{
\includegraphics[width=0.4\textwidth]{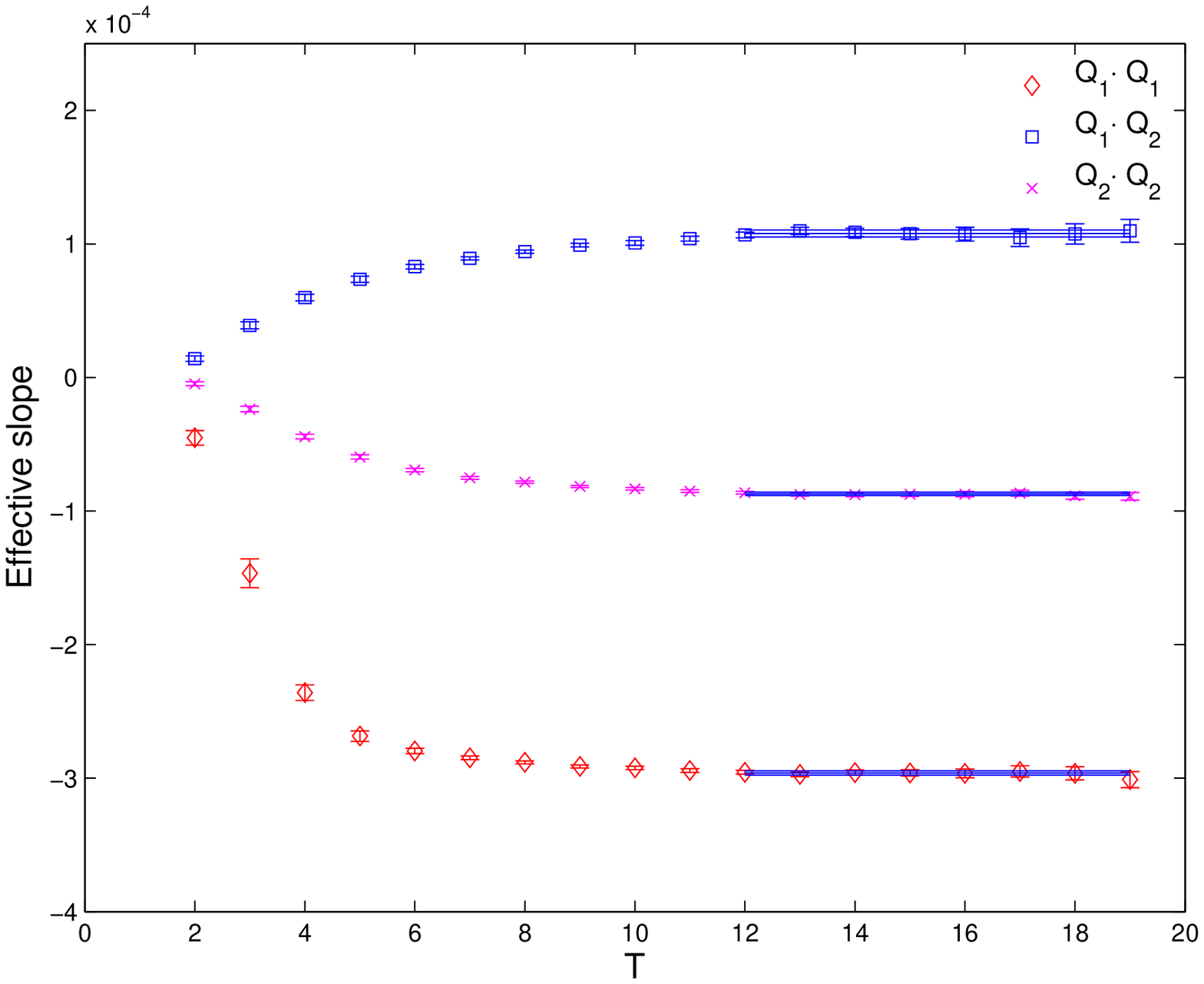} 
	\label{fig:effslope_v2_type12}
}
	\caption{Results from the combination of type 1 and 2 diagrams. The left plot gives the integrated correlators and the fitting lines. The right plot gives the effective slope plots.}
	\label{fig:type12_diagrams}
\end{figure}

\begin{figure}[!htp]
	\centering
	\subfigure[Integrated correlator]{
\includegraphics[width=0.4\textwidth]{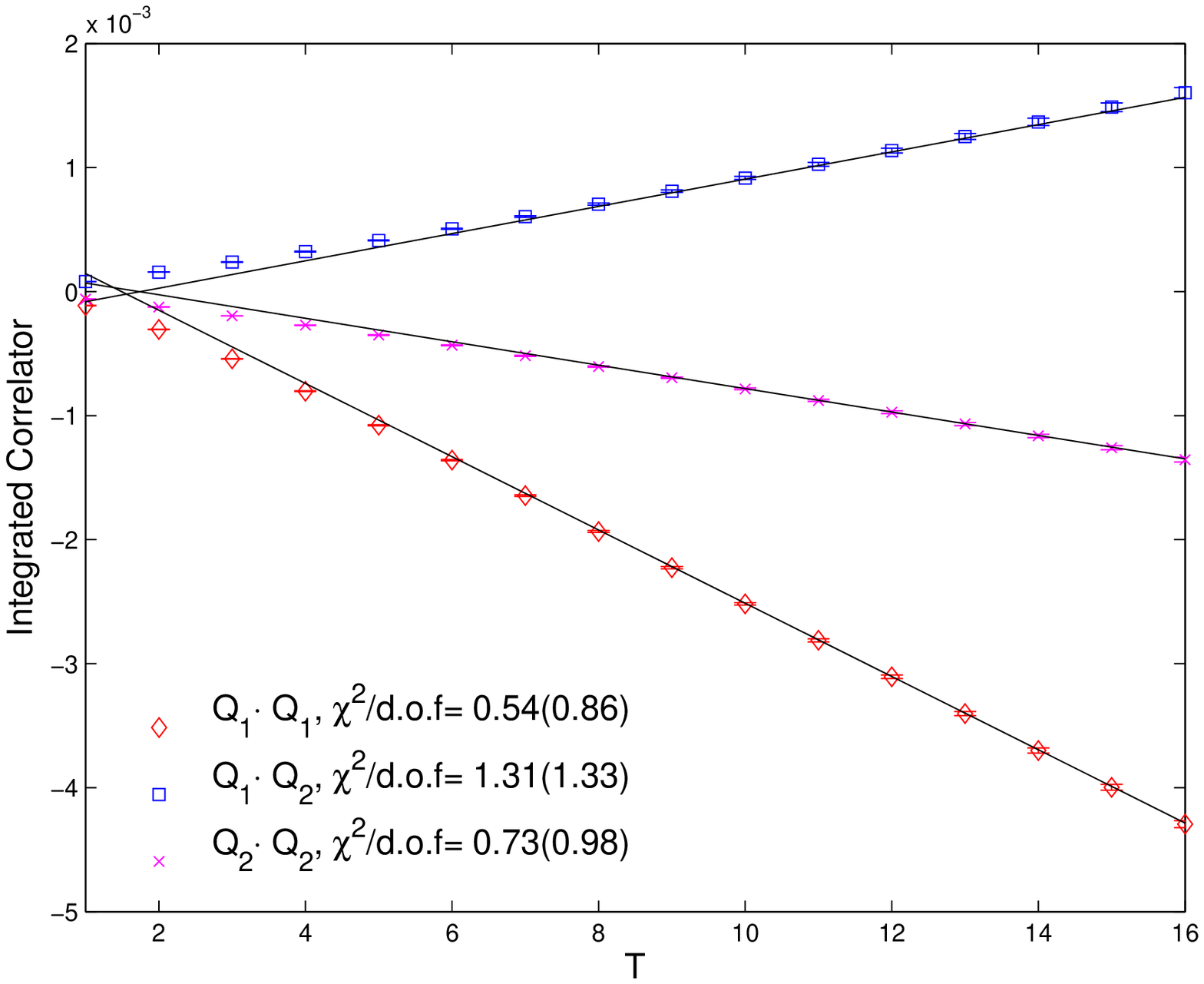} 
	\label{fig:int_corr_type123}
}
\subfigure[Effective slope]{
\includegraphics[width=0.4\textwidth]{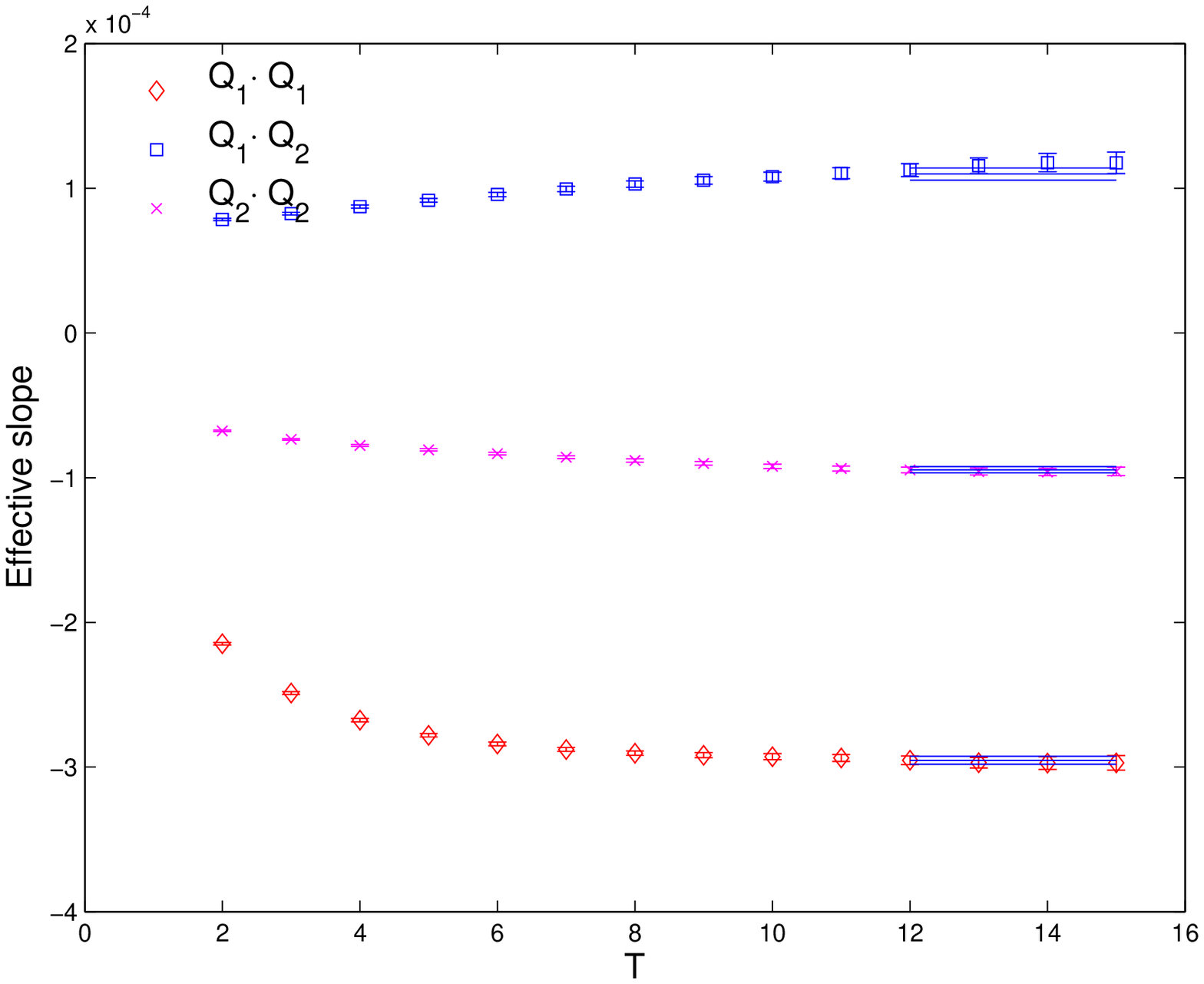} 
	\label{fig:effslope_v2_type123}
}
	\caption{Results from the combination of type 1, 2 and 3 diagrams. The left plot gives the integrated correlators and the fitting lines. The right plot gives the effective slope plots.}
	\label{fig:type123_diagrams}
\end{figure}

In our previous work, only the first two types of diagrams are included in the calculation. Now we have the data for all the diagrams, it is interesting to investigate the contribution from type 3 and type 4 diagrams. In Fig.~\ref{fig:type12_diagrams}, we give the integrated correlators and effective slopes from the combination of type 1 and type 2 diagrams. The results shown in Fig.~\ref{fig:type123_diagrams} are from the combination of type 1, 2 and 3 diagrams. In Table.~\ref{tab:type12}, we give the fitting results from difference combination of diagrams. Comparing these results, we can conclude that the contribution from type 3 diagrams is small and there is a large cancellation between type 4 (disconnected) diagrams and other types of digrams.

\begin{table}[!htp]
	\caption{Comparison of mass differences from different combinations of diagrams. All the numbers here are in units of $10^{-12}$ MeV.}
	\label{tab:type12}
	\centering
		\begin{tabular}{ccccc}
			\hline
			 Diagrams & $Q_1\cdot Q_1$ & $Q_1 \cdot Q_2$ & $Q_2 \cdot Q_2$ & $\Delta M_K$ \\
			\hline
	Type 1,2		& 1.485(8) &  1.567(38) &  3.678(56) &  6.730(96) \\
	Type 1,2,3		& 1.481(14) &  1.598(61) &  3.986(90) &  7.07(15) \\
			All & 0.754(42) & -0.16(15) & 2.70(18) & 3.30(34) \\
			\hline
 \end{tabular}
\end{table}

\section{Conclusions and outlook}
We have done a first full lattice calculation of $\Delta M_K$ with a 330 MeV pion mass, a 575 MeV kaon mass and a 949 MeV quenched charm quark mass. Our results is:
\begin{equation}
	\Delta M_K = 3.30(34)\times 10^{-12} \quad\text{MeV}
\end{equation}
Only statistical error is included here. Our result agrees very well with experimental value $3.483(6)\times 10^{-12}$ MeV. However, since we are not using physical kinematics, this nice agreement here is not extremely meaningfule. 

To perform a full calculation with physical kinematics, two difficulties must be overcome. First, we need to perform the calculation on a dynamical four flavor lattice ensemble with a smaller lattice spacing. Thus the quench effect and discretization error of charm quark can be under control. A more challenging problem is the finite volume corrections related with two pions states. This problem will become important if two pion mass is lower than kaon mass. In that case, $\Delta M_K$ in continuum limit is given by the principal part of the integral over the two pion momenta, which is quite different from a finite volume sum. A generalization of the Lellouch-Luscher method has been proposed to correct this potentially large finite volume effect~\cite{Christ:2010zz}. G-parity boundary condition is required to implement this method~\cite{Kelly:2012eh}. In summary, a full calculation of $\Delta M_K$ should be accessible to lattice QCD with controlled systematic errors within a few years. 

The author thank very much all my colleagues in the RBC and UKQCD collaborations for valuable discussions and suggestions. Especially thanks to Prof. Norman Christ for detailed instructions and discussions.

\bibliography{citation}

\providecommand{\href}[2]{#2}\begingroup\raggedright\begin{thebibliography}{1}

\bibitem{Nakamura:2010zzi}
{\bf Particle Data Group} Collaboration, K.~Nakamura {\em et~al.}, {\it {Review
  of particle physics}},  {\em J.Phys.} {\bf G37} (2010) 075021.

\bibitem{Brod:2011ty}
J.~Brod and M.~Gorbahn, {\it {Next-to-Next-to-Leading-Order Charm-Quark
  Contribution to the CP Violation Parameter $\epsilon_K$ and $\Delta M_K$}},
  {\em Phys.Rev.Lett.} {\bf 108} (2012) 121801
  [\href{http://arXiv.org/abs/1108.2036}{{\tt arXiv:1108.2036 [hep-ph]}}].

\bibitem{Donoghue:1983hi}
J.~F. Donoghue, E.~Golowich and B.~R. Holstein, {\it {LONG DISTANCE CHIRAL
  CONTRIBUTIONS TO THE K(L) K(S) MASS DIFFERENCE}},  {\em Phys.Lett.} {\bf
  B135} (1984) 481.

\bibitem{Christ:2010zz}
{\bf RBC and UKQCD Collaborations} Collaboration, N.~H. Christ, {\it {Computing
  the long-distance contribution to second order weak amplitudes}},  {\em PoS}
  {\bf LATTICE2010} (2010) 300.

\bibitem{Christ:2012np}
N.~H. Christ, {\it {Computing the long-distance contribution to the kaon mixing
  parameter $\epsilon_K$}},  {\em PoS} {\bf LATTICE2011} (2011) 277
  [\href{http://arXiv.org/abs/1201.2065}{{\tt arXiv:1201.2065 [hep-lat]}}].

\bibitem{Christ:2012se}
N.~Christ, T.~Izubuchi, C.~Sachrajda, A.~Soni and J.~Yu, {\it {Long distance
  contribution to the KL-KS mass difference}},
  \href{http://arXiv.org/abs/1212.5931}{{\tt arXiv:1212.5931 [hep-lat]}}.

\bibitem{Buchalla:1995vs}
G.~Buchalla, A.~J. Buras and M.~E. Lautenbacher, {\it {Weak decays beyond
  leading logarithms}},  {\em Rev.Mod.Phys.} {\bf 68} (1996) 1125--1144
  [\href{http://arXiv.org/abs/hep-ph/9512380}{{\tt arXiv:hep-ph/9512380
  [hep-ph]}}].

\bibitem{Martinelli:1994ty}
G.~Martinelli, C.~Pittori, C.~T. Sachrajda, M.~Testa and A.~Vladikas, {\it {A
  General method for nonperturbative renormalization of lattice operators}},
  {\em Nucl.Phys.} {\bf B445} (1995) 81--108
  [\href{http://arXiv.org/abs/hep-lat/9411010}{{\tt arXiv:hep-lat/9411010
  [hep-lat]}}].

\bibitem{Kelly:2012eh}
{\bf RBC Collaboration, UKQCD Collaboration} Collaboration, C.~Kelly, {\it
  {Progress towards $\Delta I$ = 1/2 $K \rightarrow \pi \pi$ decays with
  G-parity boundary conditions}},  {\em PoS} {\bf LATTICE2012} (2012) 130.

\end{thebibliography}\endgroup
\bibliographystyle{JHEP}

\end{document}